%% file: 4a1.tex
\documentstyle[epsf,psfig]{mn}     
\newcommand{\aeta}{A\&A}

\newcommand{\Vaco}{Variance-covariance}
\newcommand{\vaco}{variance-covariance}
\newcommand{\beq}{\begin{equation}}
\newcommand{\eeq}{\end{equation}}
\newcommand{\set}[2]{\setx{#1\,|\,#2}}  
\newcommand{\setx}[1]{\{#1\}}  
\newcommand{\bm}[1]{\mbox{\boldmath $#1$}}
\newcommand{\bM}[1]{\mbox{\textbf{\textsf{#1}}}}
\newcommand{\dotB}{\dot{\bM{B}}}
\newcommand{\parl}{\Bigl(\!\!}
\newcommand{\parr}{\!\!\Bigl)}
\newcommand{\e}[1]{\langle#1\rangle}

\newcommand{\diag}{\mbox{\rm diag}}
\newcommand{\eqw}{equivalent width}
\newcommand{\eqws}{\eqw s}

\newcommand{\syn}{\mbox{\rm\scriptsize syn}}
\newcommand{\obs}{\mbox{\rm\scriptsize obs}}

\newcommand{\paperI}{paper~I}
\newcommand{\paperII}{paper~II}

\newcommand{\Wsyn}{\mbox{$W_{\syn}$}}
\newcommand{\Wsynj}{\mbox{$W_{\syn\,j}$}}
\newcommand{\Isynj}{\mbox{$I_{\syn\,j}$}}
\newcommand{\Wobs}{\mbox{$W_{\obs}$}}
\newcommand{\Wobsj}{\mbox{$W_{\obs\,j}$}}

\newcommand{\ns}{n_{\star}}
\newcommand{\nss}{n_{\mbox{\scriptsize s}}}

\newcommand{\nl}{{n_{\lambda}}}

\newcommand{\dfrac}[2]{\displaystyle \frac{#1}{#2}}
\newcommand{\sumi}{\sum_{i=1}^{\ns}} 

\begin{document}
 
   \title
[Error analysis.]
{Error analysis for stellar population synthesis as an inverse problem.}
               
   \author
    [J. Moultaka, D. Pelat]
    {J. Moultaka, D. Pelat \\  
     DAEC, Unit\'e associ\'ee au CNRS et \`a l'Universit\'e Denis Diderot,
     Observatoire de Paris, section de Meudon\\
     F--92195 Meudon Cedex, France\\
     E-mail: didier.pelat@obspm.fr}

 \date{Received : to be inserted later; accepted : to be inserted later}
\maketitle

\begin{abstract}
 Stellar population synthesis can be approached as an inverse problem.
 The physical information is extracted from the observations through an inverse model.
 The process requires the transformation of the observational errors into model errors.
 A description is given for the error analysis to obtain objectively the errors in the model.

 Finding a solution for overdetermined and under-determined case was the purpose of two preceding papers.
 This new one completes the problem of stellar populations synthesis by means of a data base, by providing practical formul\ae\ defining the set of acceptable solutions.
 All solutions within this set are compatible, at a given confidence level, with the observations.  
\end{abstract}

\begin{keywords}
  Galaxies: stellar content
  -- methods: data analysis
  -- methods: numerical
\end{keywords}
%
 
\section{Introduction}
 Important astrophysical issues could be solved if we were able to deduce the stellar population from the integrated light received from a far-away galaxy (or a region of a galaxy).
 Two methods try to solve this problem : the {\em stellar data base synthesis} and the {\em evolutionary population synthesis}.
 The first method is a more empirical approach, it relies strictly on the spectral data base of stars or stellar aggregates, while the second one is a more theoretical approach; it relies on our best knowledge on the formation and evolution  of stars and stellar aggregates.
 Stellar data base synthesis is used in both inverse and forward modeling, while evolutionary  population synthesis is mostly applied as a forward model.
 In our opinion the two methods are complementary and should be used jointly in an effort to solve this difficult problem.
 Both methods are subject to errors and ideally their results should always be presented with a {\em set} of acceptable solutions rather than just {\em a} solution.
 The purpose of this paper is to establish this set of acceptable solutions for the {\em stellar data base synthesis}.

Pelat (1997 and 1998; hereafter \paperI\ and \paperII) re-investigated the stellar data base synthesis method and proposed new and fast algorithms to search for physical solutions.
 The sensitivity of this solution to the observational errors was discussed in general terms: Monte-Carlo simulations were performed but no real error analysis had been done.
 This paper provides the user with {\em practical} formul\ae\ which give an estimate of the domain of acceptable solutions around the solution. 

 In the following \S~\ref{ch.02}, we recall briefly the solution we have proposed and then we focus in great detail on the question of error analysis.

\section{Population synthesis as an inverse problem}\label{ch.02}

 First we specify what is meant by {observables} and {parameters} in the description of the physical model.

 {\em Observables.}
It is usually agreed to adopt \eqws\ as spectral observables, in order to avoid data reduction errors and to minimize extinction problems.
 These observables are noted here as the vector $\bm{\Wobs}$.
 The aim of the stellar population synthesis is to find the combinations of stars which best reproduce the \eqws\ $\bm{\Wobs}$ of the observed object (say, a galaxy).

{\em Model parameters.}
 The galaxy is described by the unknowns $k_i$. They give the proportional contribution to the luminosity due to stars of type $i$ at a reference wavelength $\lambda_0$.
 The assembly of $k_i$s is described by the vector $\bm{k}$.

 The equation governing the stellar population synthesis problem is a non linear one, where a set of synthetic \eqws\ is a function of a set of stellar luminosity contributions $k_i$.
 We have:
\beq\label{eq.1}
\Wsynj = \frac{\sumi W_{ji}I_{ji}k_i}{\sumi I_{ji}k_i}\,,\quad
j=1,\ldots,\nl\,.
\eeq
 Where, $\nl$ is the number of observed \eqws; $\ns$ is the number of stars used for the synthesis; $W_{ji}$ and $I_{ji}$ are respectively the \eqw\ and  continuum flux at wavelength $\lambda_j$ of stars of kind $i$.
 The continuum fluxes are normalized to one at the reference wavelength $\lambda_0$.
 For short we write Eq.~(\ref{eq.1}): $\bm{W_{\syn}}=\varphi(\bm{k})$. 

 In addition we have the condition that the contribution of a star to the luminosity cannot be negative and that the normalized proportions $k_i$ add up to one. A physical solution of Eq.(\ref{eq.1}) therefore lies in the vector set:
\beq\label{eq.setS}
{\cal S}=\set{(k_1,\ldots,k_{\ns})}{k_i\ge 0, {\textstyle \sumi} k_i=1}.
\eeq
 This set ${\cal S}$ is a simplex (e.g. an equilateral triangle for a data base of three stars, a tetrahedron for a data base of four and so on). 
 The set of all \eqws\ which are able to be exactly synthesized by the data base is the image of ${\cal S}$ by $\varphi$.
 This is the synthetic domain ${\cal W}$ introduced in \paperI:
\beq\label{eq.setW}
{\cal W}=\set{\bm{W}}{\bm{W}=\varphi(\bm{k}), \forall \bm{k}\in{\cal S}}.
\eeq
 If $\bm{\Wobs}\in{\cal W}$ there is at least one exact solution $\bm{k}$ such that $\varphi(\bm{k})=\bm{\Wobs}$.
 
 The set ${\cal K}$ of all exact solutions is given formally by: ${\cal K}=\varphi^{-1}(\bm{W}_{\obs})$. This set may be empty.
 The solution set ${\cal K}$ is non-empty if there exists at least one vector $\bm{k}$ solution  of the following system:
\beq\label{eq.Bk}
\left.
\begin{array}{r}
\bm{k}\ge 0\,,\\[0.8ex]
\bM{A}\bm{k}=\bm{0}\,,\\[0.8ex]
\bm{b}^T\bm{k}=1\,.
\end{array}
\right\}
\eeq
 The matrix elements of $\bM{A}$ are : ${[\bM{A}]}_{ji}=(W_{\obs j}-W_{ji})I_{ji}$ and $\bm{b}^T$ is a line of `ones'.
 This system may or may not  possess a solution.

\subsection{Test for the existence of an exact solution}
 It is very easy to check if the above system~(\ref{eq.Bk}) possesses at least one solution.
If we add to it a linear equation to be maximized  (e.g. $\bm{c}^T\bm{k}=\max!$), this system turns into a linear program.
 In fact, it is not necessary to explicit that supplementary linear equation, we only have to solve the linear program for a {\em feasible} solution (see for example Press et al. 1992 section 10.8).
 If no feasible solution exists this means that the galaxy cannot be exactly synthesized by the data base. 
 This test is extremely rapid.

\subsection{Under-determined case: $\nl \le \ns - 1$ }

If the galaxy can be exactly synthesized (${\cal K}$ is non-empty), it has been demonstrated in \paperII\ that ${\cal K}$ is a polytope, i.e. it is the convex hull of some finite extreme solutions.
 A solution $\bm{k}$ is extreme if it possesses at least $n_0-1$  components (stellar contributions) equal zero.
 The key parameter $n_0$ is the dimension of the null space associated with matrix $\bM{A}$.
  The data base is non-degenerate for the observation if we have $n_0=\ns-\nl$.
 
 If the galaxy cannot be synthesized, an approximate solution is found with a data base reduced to at most $\nl$ stars.
 The procedure is described in the next section.

\subsection{Overdetermined case: $\nl \ge \ns$ }\label{ch.1.2}

In the overdetermined case, there is most probably no exact solution of~(\ref{eq.Bk}), the observation is not on the synthetic surface and one must content oneself with an approximate solution.
 This approximate solution $\bm{k}$ is usually accepted in the least square meaning, that is $\bm{k}$ must minimize a quadratic form :
\begin{equation}\label{eq.5}
 D^2 =
 (\bm{W}_{\obs}-\varphi(\bm{k}))^T\bm{\Sigma}^{-1}
 (\bm{W}_{\obs}-\varphi(\bm{k}))\,,
\end{equation}
where $\bm{\Sigma^{-1}}$ is a positive definite `weight' matrix.
 It has been shown in \paperI\ that a very successful estimate of $\bm{k}$ is found near the `first-guess' $\bm{k}_0$:
\[
 \bm{k}_0 = 
(\bM{B}^T\bM{B})^{-1}\bm{b}[\bm{b}^T(\bM{B}^T\bM{B})^{-1}\bm{b}]^{-1}\,,
\]
where $\bM{B}$ is equal to the matrix $\bM{A}$ augmented with the line $\bm{b}^T$.
 It is shown in appendix~\ref{ap.1} how one can get rid of the positivity constraint $\bm{k}\ge \bm{0}$ when searching for a minimum in the neighbourhood of $\bm{k}_0$.
 The estimate $\bm{k}_0$ is the unique solution of the problem if the matrix $\bm{\Sigma}^{-1}$ is diagonal with diagonal elements equal to $I^2_{\syn\,j}$ ($I_{\syn\,j}=\sum_{i=1}^{\ns}I_{ji}k_i$).
 It was argued in \paperI\ that a reasonable weight matrix should not be very different from this particular diagonal matrix.

\section{Purposes of the error analysis}

 An observation of a galaxy is not just the point $\bm{W}_{\obs}$ in $W$-space, it includes all points interior to an error zone around the observation.
 If we assume the observational errors to be Gaussian, this error zone is an hyper-ellipsoid around $\bm{W}_{\obs}$.
 We have observed $\bm{W}_{\obs}$ but  we consider that we could have observed, with probability $\gamma$, any other point at the interior of this hyper-ellipsoid.
 This set of `probable' points: ${\cal E}_\gamma$, is formally defined as
\beq\label{eq.setE}
{\cal E}_\gamma =
\set{\bm{W}}{(\bm{W}-\bm{\Wobs})^T\bM{V}^{-1}(\bm{W}-\bm{\Wobs})\leq 
F_{\chi^2}^{-1}(\gamma)}\,,
\eeq
where $\bM{V}$ is the \vaco\ matrix of the observation and $F_{\chi^2}$ is the distribution function of a $\chi^2$ variate with $\nl$ degrees of freedom.
 For short, we call ${\cal E}_\gamma$ the `error ellipsoid'.
 We recall that $\bM{V}$ is equal to $\e{\Delta\bm{W}\Delta\bm{W}^T}$, where $\e{\hspace{1em}}$ stands for the expectation and $\Delta\bm{W}$ for a variation of $\bm{W}$ around its mean.

 The purpose of the error analysis is to determine how the observational errors are transformed by the inversion process.
 Usually one tries to solve two problems as follows: i) deduce $\bM{V}_k$ the \vaco\ matrix of the solution knowing the matrix $\bM{V}$ for the observational errors; ii) construct ${\cal K}_\gamma$, the set of all solutions that fall within the `error ellipsoid', i.e. ${\cal K}_\gamma=\varphi^{-1}({\cal E}_\gamma)$.

 These problems are simplified if the error ellipsoid corresponds to small deviation $d\bm{W}$ around the observation.
 In that approximation one can linearize $\varphi^{-1}$, which implies that ${\cal K}_\gamma$ is also an hyper-ellipsoid.
 We have:
\beq\label{eq.setK}
{\cal K}_\gamma \approx
 \set{\bm{k}'}
     {(\bm{k}'-\bm{k})^T\bM{P}(\bm{k}'-\bm{k})\le F_{\chi^2}^{-1}(\gamma)}\,.
\eeq
The matrix $\bM{P}$ would be equal to $\bM{V}_k^{-1}$ if $\bM{V}_k$ were invertible but this is not the case here: the matrix $\bM{V}_k$ is singular because of the normalization constraint on $\bm{k}$.
 Using the definition of ${\cal K}_\gamma$ (\ref{eq.setK}), one is able to test if an alternative solution $\bm{k}'$ is acceptable at the confidence level $\gamma$. 

 The results of the error analysis are given in \S~\ref{ch.3}.
 Before that, we give in \S~\ref{ch.15} an overview of the main processes at the origin of the observational errors and in \S~\ref{ch.2} we introduce information content held by an observation with regard to the synthesis problem we want to solve.
  
\subsection{Origin of the errors}\label{ch.15}

 Errors on \eqws\ from spectra of galaxies have several origins.
 The most important one is induced by the empirical process of continuum intensity plotting. This leads to a wide uncertainty on \eqws\ estimates, caused by the fact that the continuum intensity lies on the denominator in the \eqw's expression.
 Contributing to the uncertainties in this empirical approach are the blending of the absorption lines plus the possible presence of emission lines as in active galactic nuclei spectra.
 The limited wavelength range of the study plays also a crucial role in this domain.

 Another phenomenon in active galactic nuclei spectra is the presence of an additional non stellar continuum which leads to the dilution of lines, i.e. the reduction of \eqws\ in comparison with those of normal galaxies.
 The effect of this non stellar continuum can be estimated and \eqws\ can be corrected; but obviously, this estimation is also subject to errors.
 Finally, velocity dispersion in the galaxies broadens lines and is also a source of errors.
 Considering these facts, typical errors of $10\%$ are present in the \eqws\ of strong lines.

 We did not take into account the errors in the equivalent width of the lines in the data base itself because they are supposed to be negligible compared with errors in the equivalent width of the lines from the galaxy.
 However we provide the user in \S~\ref{ap.213} with a test able to validate this hypothesis. 
 If these errors were not negligible, one should not perform the synthesis. 

\section{Information content of an observation.}\label{ch.2}

 It is clear that if the synthetic domain ${\cal W}$ is entirely contained within the observational error ${\cal E}_\gamma$, the observation cannot discriminate, at the level $\gamma$, between the different stellar populations.
 All possible models which fall within ${\cal E}_\gamma$ must be considered as indistinguishable.
 In other words, the observation bears not enough information to differentiate between various possible stellar populations of the galaxy.
 The size of the error ellipsoid ${\cal E}_\gamma$ is in some way the resolution at which we try to sort between the different models.

 On the contrary, if the observation is so good that one can consider the `error-bars' to be null, the observation indeed convey a maximum information: $\bm{\Wobs}$ {\em must} be synthesized by the data base.
 Under this extreme hypothesis, ${\cal E}_\gamma$ is reduced to the single point $\setx{\bm{\Wobs}}$ which has a zero measure in ${\cal W}$.

 Following a suggestion of R. Barrett we define the information $I_\gamma$ brought by the observation as being a function of the measure of ${\cal E}_\gamma$ within ${\cal W}$.
 More precisely $I_\gamma$ will be a function of the probability $p_\gamma$ that the image by $\varphi$ of a point drawn at random from the {\em a priori} distribution of the stellar populations falls within the error ellipsoid.
 That is:
\beq\label{eq.info}
p_\gamma = \int\limits_{{\cal E}_\gamma \cap {\cal W}}
             \!\!\pi(\bm{w})\,d\bm{w}\,,\quad
\bm{w}=\varphi(\bm{k})\,,
\eeq
where $\pi$ is the {\em a priori} probability density function of the \eqws.
 We shall evaluate this $\nl$-dimensional integral by a Monte-Carlo method (see appendix~\ref{ap.1}).

 The probability $p_\gamma$ may be zero under two circumstances: (i) $\bm{\Wobs}$ is within the synthetic domain and ${\cal E}_\gamma$ is reduced to $\setx{\bm{\Wobs}}$ only; (ii) the observation and its associated error domain are outside the synthetic domain: ${\cal E}_\gamma \cap {\cal W}=\emptyset$.
 In each case we have maximum information of different nature conveyed by the observation: (i) we {\em know} that the true stellar population lies in the solution set (provided that all stars present in the galaxy are also present in the data base); (ii) we {\em know} that the data base is incomplete, more stars must be added. 

 In a first approach, it seems reasonable to define the information brought by an observation by: $I_\gamma=1-p_\gamma$, i.e. it is the probability that a stellar population drawn at random from the simplex ${\cal S}$ induces \eqws\ that do {\em not} fall within the error domain.
 By `at random' we mean according to $\pi$, the {\em a priori} distribution of the \eqws.
 In practice we use the transform by $\varphi$ of an {\em a priori} distribution of the stellar populations which is uniform over ${\cal S}$.
 Defined like this, $I_\gamma$ is zero if all models are validated by the observation (no information) and one if the solution set has a zero probability to be reached by a random model (maximum information, one cannot get there at random).

 An illustration of the information content within an observation is given in Fig~\ref{fig.1}, see also Fig.~4 of \paperII.
 Here, at the 1-$\sigma$ confidence level ($\gamma=0.683$) the information content is $I_\gamma=0.92$.
 This observation has therefore a high information content.
 It is located on a part of the synthetic domain, which is seldom reached by a random synthesis following an {\em a priori} uniform distribution in $\bm{k}$.
  
\section{Error analysis in practice.}
\label{ch.3}
 In this section, we limit ourselves to the practical results of the error analysis.
 We refer to appendix B for all the computations needed to derive the various matrices which appear below.

 The key parameter here is the number $\nss$ of stars necessary to define the solution (or the extreme solutions).
 In the overdetermined case $\nss=\ns$, while in the under-determined case $\nss=\nl+1$ if the galaxy can be synthesized and $\nss=\nl$ if it is not.
 It is, in both under- and overdetermined cases, less or equal to $\nl+1$.
 We have a {\em regular} case if $\nss=\nl+1$ and a {\em singular} one if $\nss<\nl+1$.
 We consider separately these two cases. 
 In both cases, $\bM{V}$ stands for the \vaco\ matrix of the observations and $\bm{s}$ for the list of the indices of the stars retained in the solution.

\subsection{ Regular case $\nss=\nl+1$.}
 Note that in this case the galaxy belongs to the synthetic domain. 
 A solution $\bm{k}$ is any barycentric combination of $n_K$ extreme solutions  denoted by $\bm{k}_s$ ($n_K$ may be equal to one).
 At least $\ns-(\nl+1)$ components of $\bm{k}_s$ are equal to zero.
 The components of $\bm{k}_s$ are in addition subject to errors, due to the presence of errors in the data.
 In the tangent approximation (small errors), the \vaco\ matrix of the extreme solutions is a singular matrix given by:
\begin{equation}
\bM{V}_{k_s} =
 \bM{K}\parl
 \begin{array}{lc}\bM{V}&\hspace{-2ex}\bm{0}\\
\bm{0}^T&\hspace{-2ex}0\end{array}
 \parr\bM{K}^T\,,\quad
 \bM{K}=\bM{W}^{-1}\,.
\end{equation}
 The elements of the matrix $\bM{W}$ are:
\begin{equation}
\begin{array}{l@{\;=\;}ll}
[\bM{W}]_{ji} & (\Wobsj-W_{js(i)})\dfrac{I_{js(i)}}{\Isynj}\,,\quad&
j=1,\ldots,n_s-1\,;\\
 & 1\,,& j=n_s\,,
\end{array}
\end{equation}
where $\bm{s}(i),\,i=1,\ldots,\nss$ run over the stars retained in the solution.
 The set of acceptable extreme solutions is an ellipsoid centered around the extreme solution $\bm{k}_s$.
 Its characteristic matrix $\bM{P}_s$ is given by:
\begin{equation}
\bM{P}_s = 
 \bM{W}^T\parl
 \begin{array}{lc}\bM{V}^{-1}&\hspace{-2ex}\bm{0}\\
\bm{0}^T&\hspace{-2ex}0\end{array}
 \parr\bM{W}\,,
\quad
\bM{W}=\bM{J}^{-1}\bM{B}\,,
\end{equation}
where $\bM{J}$ is a $\nl+1$ by $\nl+1$ diagonal matrix:
\[
\begin{array}{r@{\,=\,}ll}
 {[\bM{J}]}_{jj} & \Isynj\,,\quad &j=1,\ldots,\nl\,;\\
{[\bM{J}]}_{jj} & 1\,,\quad &j=\nl+1\,.
\end{array}
\]
The set of acceptable solutions may be considered as the convex hull of all these acceptable extreme solutions.

\begin{table}
\caption
{The extreme solutions and errors associated with the observation and data base illustrated in Fig.~\ref{fig.1}.
 Solution number $s$ is $\bm{k}_s$; $\sigma\bm{k}_s$ is the standard deviation vector associated with the solution.
 For the extreme solution Nr.1 we also give $\Delta=2\sigma c_\gamma$ in order to allow a direct comparison with Fig.~\ref{fig.1}.
 The parameter $c_\gamma$ determines the size of the error ellipsoids.
 We have: $c_\gamma=[F^{-1}_{\chi^2}(\gamma)]^{1/2}$, e.g. with $\gamma\approx 0.683$ and $\nl=2$ degrees of freedom one gets $c_\gamma\approx 1.515$.
 The quantity $\Delta$ is the difference between the maximum and the minimum attained by the stellar contributions of an observation which is constrained to move on the error ellipsoid.
 In spite of the non-linear nature of the problem illustrated here, it is clear from Fig.~\ref{fig.1} that the error analysis under the tangent approximation is excellent.}
\label{tab.1}
\begin{tabular}{rrrrrr}
\hline
\multicolumn{1}{r}{ } &
\multicolumn{1}{c}{$k_1$} &
\multicolumn{1}{c}{$k_2$} &
\multicolumn{1}{c}{$k_3$} &
\multicolumn{1}{c}{$k_4$} &
\multicolumn{1}{c}{$k_5$} \\
\hline
$\bm{k}_1$ & 0.306 & 0.245 & 0.000 & 0.449 & 0.000 \\
$\sigma\bm{k}_1$ &
$\pm$0.115 & $\pm$0.094 & & $\pm$0.119 & \\
$\Delta$ &
0.347 & 0.286 & & 0.361 & \\[0.8ex]

$\bm{k}_2$ & 0.491 & 0.000 & 0.218 & 0.291 & 0.000 \\
$\sigma\bm{k}_2$ &
$\pm$0.094 & &  $\pm$0.111 &  $\pm$0.124 & \\[0.8ex]

$\bm{k}_3$ & 0.231 & 0.000 & 0.000 & 0.308 & 0.462 \\
$\sigma\bm{k}_3$ &
$\pm$0.111 & & &  $\pm$0.115 & $\pm$0.150\\
\hline
\end{tabular}
\end{table}

\begin{figure}
\centering
\psfig{figure=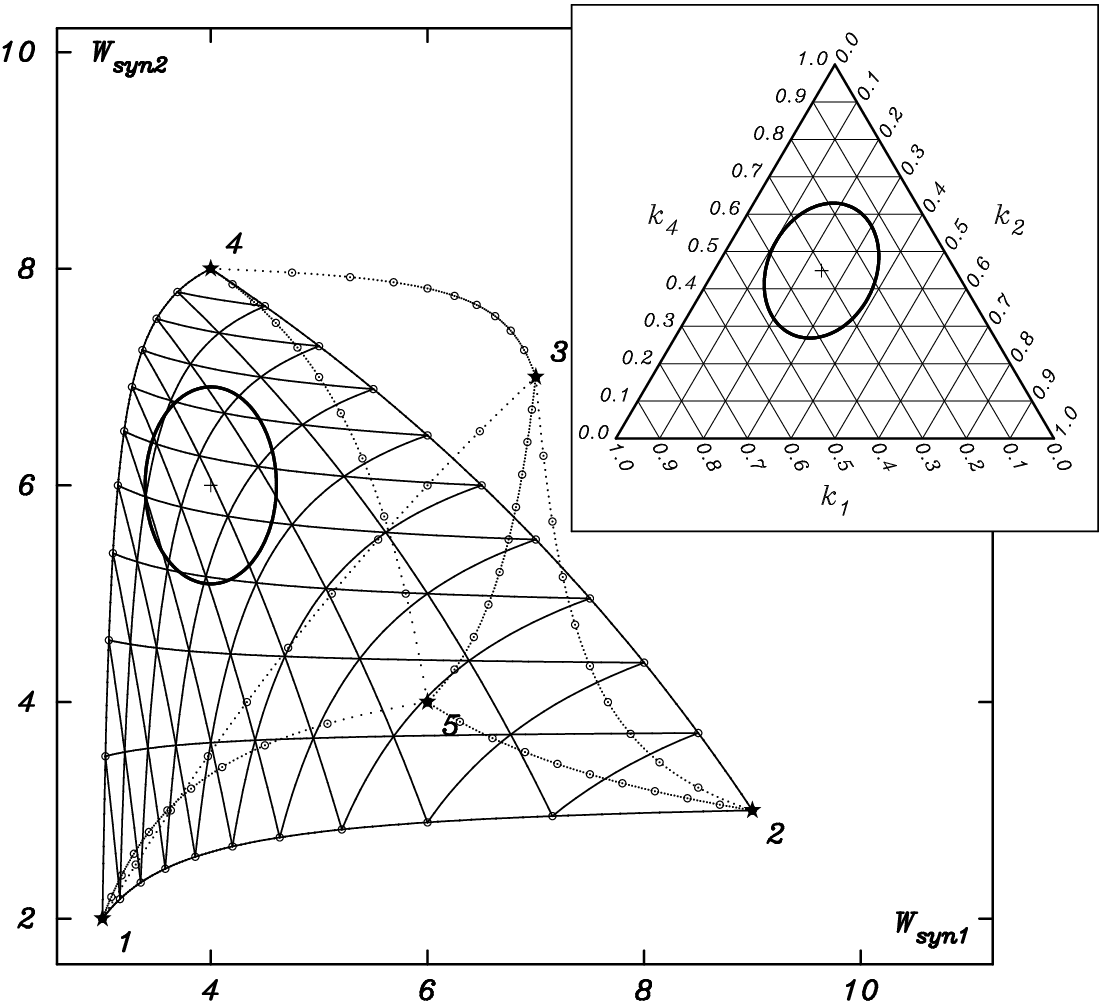,height=7.7cm}
\caption
{Illustration of the error analysis for the extreme solution $\bm{k}_1$.
Around the observation $\bm{\Wobs}=(4,6)$ there is an error ellipsoid defined by a \vaco\ matrix $\bM{V}$.
 Here this matrix is diagonal, the square root of its diagonal elements (the standard deviations of the observations) are set to $10\%$ of $\bm{\Wobs}$.
 The synthetic domain is defined by a data base of 5 stars identical to the one given in Table~1 of \paperII. 
 This observation and its error ellipsoid contain the information $I_\gamma\approx 0.92$ at the 1-$\sigma$ confidence level: $\gamma=0.683$.}
\label{fig.1}
\end{figure}
 
 We have illustrated in Fig.~\ref{fig.1} the error analysis performed on an observation subject to errors of $10\%$.
 The population synthesis was done with the same data base used in \paperII. The results are given in Table~\ref{tab.1}.
 The $\bm{k}_s,\,s=1,2,3$ are the three extreme solutions; $\sigma\bm{k}_s$ is the standard deviation vector associated with the solution (it is the square root of the diagonal of $\bM{V}_{k_s}$); $\Delta$ is the extreme range attained by the $k_i$s by a point, which is constrained to move on the error ellipsoid.
 The analysis was done at the confidence level of 1-$\sigma$, that is: $\gamma\approx 0.683$.

\subsubsection{Merit order among extreme solutions}\label{ch.511}
 The product of the non-zero eigenvalues of $\bM{V}_k$ is proportional to the surface of the ellipsoid ${\cal K}_\gamma$.
 This quantity allows to sort the extreme solutions in order of merit, from the smallest surface (highest merit) to the greatest surface (smallest merit).
 In a synthesis with many extreme solutions it is advisable to retain only the solutions which possess the highest merit. 

\subsection{ Singular case $\nss<\nl+1$.}
 In the singular case, an approximate solution is searched by minimizing a distance from the observation to the synthetic domain.
 This distance is usually the elliptical distance $D$, defined by Eq.~(\ref{eq.5}), and it requires a positive definite matrix $\bm{\Sigma}$.  
 However, we suppose below that $D$ is the Euclidean distance through the variable transformation described in appendix~\ref{ch.B26}.
 The ellipse has been transformed into the unit circle and $\bm{\Sigma}=\bM{I}$.

 The error analysis is complicated by the fact that one must perform a projection on a plane which is tangent to the synthetic domain around the approximate solution.
 We describe step by step the operations which lead to the result.
\begin{description}
\item [{\bf Step $A$}:]
 First, construct the $\nl$ by $\nss$ matrix $\dot{\bM{A}}$ of elements: $[\dot{\bM{A}}]_{js(i)}=(\Wsynj-W_{js(i)})I_{js(i)}$.
 The index $s(i)$ runs over the $\nss$ stars defining the solution.
 We need also the diagonal matrix $\bM{J}_A=\diag(I_{\syn 1},\ldots,I_{\syn\nl})$.
\smallskip
 
\item [{\bf Step $Q$}:]
 The matrix $\bM{G}=\bM{J}_A^{-1}\dot{\bM{A}}$ is constructed.
 A rank deficiency QR algorithm is performed on $\bM{G}$ (e.g. using {\tt dgeqpf} from lapack).
 Following this operation one gets: $\bM{G}\bm{\Pi}=\bM{Q}\bM{R}$.
 The matrix $\bM{Q}$ is a $\nl$ by $\nl$ square matrix partitioned into two blocks: $\bM{Q}=(\bM{p}|\bM{q})$.
 The first block is formed by the $\nss-1$ first columns of $\bM{Q}$.
 An orthogonal projector $\bM{H}=\bM{pp}^T$ is constructed.
\smallskip

\item [{\bf Step $V$}:]
 We define the matrix $\dotB$ equal to $\dot{\bM{A}}$ with a supplementary line of `ones' and the matrix $\bM{K}=(\dotB^T\dotB)^{-1}\dotB^T\bM{J}$ where $\bM{J}$ has already been defined ($\bM{J}=\diag(I_{\syn 1},\ldots,I_{\syn\nl},1)$\,).

 The \vaco\ matrix of the solution is:
\[
\bM{V}_k = \bM{K}
\parl
 \begin{array}{cc}
 \bM{H}\bM{V}\bM{H}^T&\hspace{-2ex}\bm{0}\\
 \bm{0}^T&\hspace{-2ex}0
 \end{array}
 \parr
\bM{K}^T\,.
\]

\item [{\bf Step $P$}:]
 We define the oblique projector $\bM{M}$:
\[
 \bM{M} = \bM{p}\bM{p}^T+\bM{q}\bM{q}^T\bM{V}\bM{p}
(\bM{p}^T\bM{V}\bM{p})^{-1}\bM{p}^T\,,
\]
 and compute: $\bM{W}=\bM{J}^{-1}\dotB$.
 The matrix $\bM{P}$ entering the definition of the acceptance zone in Eq.~(\ref{eq.setK}) is given by:
\[
 \bM{P} = \bM{W}^T
\parl\begin{array}{cc}
 \bM{M}^T\bM{V}^{-1}\bM{M} & \hspace{-1.0ex}\bm{0}\\
 \bm{0}^T&\hspace{-1.0ex} 0
\end{array}
\parr
\bM{W}\,,\quad\bM{W}=\bM{J}^{-1}\dotB\,.
\]
\end{description}
 Note that in the regular case we have: $\bM{K}=\bM{B}^{-1}\bM{J}$ and in the singular case we have: $\bM{K}=(\bM{B}^T\bM{B})^{-1}\bM{B}^T\bM{J}$.
 Therefore $\bM{K}$ is always the least square solution of the linear system $\bM{B}\bM{K}=\bM{J}$.
(In both cases $\bM{B}$ is constructed with \Wsyn\ and the stars contributing to the synthesis.) 

\section{The example of NGC\,3521.}
\begin{figure*}
\centering
\psfig{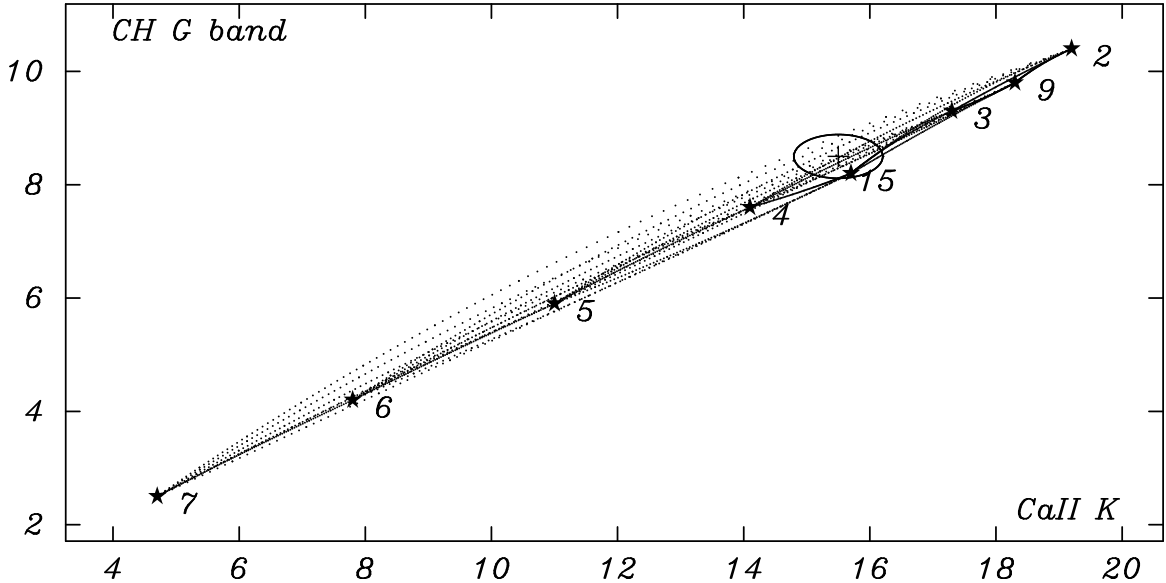}
\caption
{The synthetic domain of the 8 Bica's clusters identified by their numbers
(2--7,9,15) in the 2 dimensions \eqws\ space corresponding to the Ca\,II K line and the CH G band.
 The cross is the position of NGC\,3521 in this space and the ellipsoid corresponds to a $3\%$ standard deviation on the galactic \eqws\ at a confidence level of $\gamma=0.683$.
 At this level the information content of this observation is $I_\gamma=0.92$ (see \S\,\ref{ch.2}).
}
\label{fig.2}
\end{figure*}
 To illustrate our study we take, as an example, the clusters of Bica's data base (see Bica 1988) and its galaxy S3 (NGC\,3521).
 As our aim is to visualize the error zone, we shall use only two lines (i.e. $\nl=2$) this means two degrees of freedom for the error analysis.
 (Indeed our analysis is valid for any number of lines.)
 From the 35 clusters constituting the data base, we select according to Bica's results the eight clusters (i.e. $\ns=8$) contributing to the synthesis.
 We have chosen the two largest \eqws\ of the galaxy which correspond to the Ca\,II K line ($\lambda_K=3933$\AA) and the CH\,G band ($\lambda_G=4301$\AA); the reason of this choice is that the relative errors on the strongest lines are generally the smallest.
 Typical errors on such \eqws\ are in the order of $10\%$, but for the sake of argument, we take an error of only $3\%$.
 We also assume, that there is no correlation between 
\eqws\ errors ($\rho=0$) even if this is probably not the case in reality.

 Fig.~\ref{fig.2} shows that the galaxy can be synthesized exactly. Our method gives 16 different extreme solutions.
 Table~\ref{tab.2} shows these several extreme solutions $\bm{k}_s$ where $s=1,\ldots,16$; also shown are the standard deviations vectors $\sigma\bm{k}_s$ corresponding to 1-$\sigma$ confidence level; and the standard deviation $\sigma'\bm{k}_s$ taking into account a $1\%$ error in the data base.
 The solutions are sorted by increasing order of merit (i.e. surface$^{-1}$).

\begin{figure}
\centering
\psfig{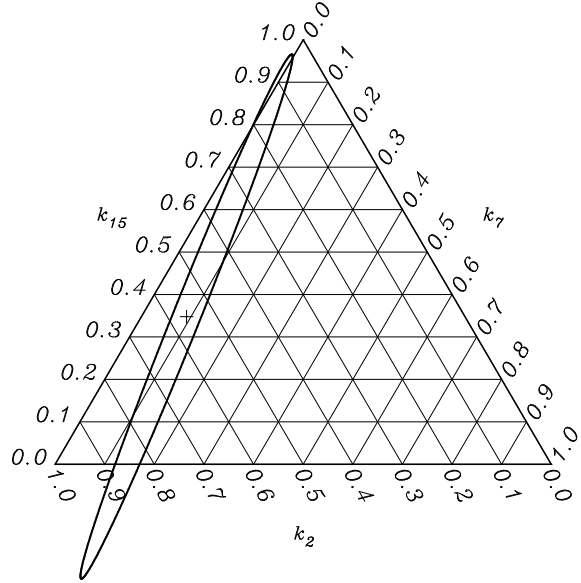}
\caption
{
The extreme solution of NGC\,3521 synthesis that corresponds to the highest merit.
 The three clusters contributing to the solution are number 2, 7 and 15 (see solution $\bm{k}_1$ in table~\ref{tab.2}).
 The ellipsoid of errors is the tangent approximation of image by application $\varphi^{-1}$ of the error ellipsoid of figure~\ref{fig.2}.
}
\label{fig.3}
\end{figure}

 Figure~\ref{fig.2} demonstrates clearly the impact of the synthetic domain structure $D$ on the solution errors.
 Indeed the clusters constituting the data base are nearly aligned in  $W$-space and their continuum intensities are such that the synthetic domain is compressed (i.e the lines joining the clusters are very close and nearly parallel to each others).
 As a consequence large errors are present in the extreme solutions (see also figure~\ref{fig.3}).
 It is important to bear this situation in mind when a data base of clusters (or stars) is chosen. Ideally the resolution at which the \eqws\ space is sampled should be adapted to the quality of the observations.
 Better observations allow a finer coverage of the $W$-space i.e. more clusters (or stars) may be added.
 We see in table~\ref{tab.2} that the contribution of data base errors to the solution errors is negligible.
 Finally, the best solution (as defined in \S \ref{ch.511}) is solution Nr.\,1; the solutions~1,~4 and 11 are the three solutions among the sixteen extreme solutions that agree well with Bica's solution (see $\bm{k}_{\small\mbox{Bica}}$ in Tab.~\ref{tab.2}).
 This result is very satisfying, because we used minimal information by considering only two \eqws.

\section{Conclusion.}
 We provide explicit formul\ae\ to perform a complete error analysis on stellar population synthesis by means of a data base.
 The hypothesis underlying the results are that errors on the data base itself are negligible compared with errors on the observed galaxy.
 The method also supposes that the errors are reasonably Gaussian and further demands the knowledge of the error matrix (the \vaco\ matrix) of the observed \eqws.

 The results provided are (i) the \vaco\ matrix of the extreme solutions (regular case) or of the approximate solution (singular case); (ii) the matrix defining a zone in the solution space where the stellar populations are indistinguishable at a confidence level $\gamma$.

 We also introduced information contained in an observation with respect to the problem to be solved.
 This information is evaluated by a Monte-Carlo method using as input the {\em a priori} distribution of the stellar population.
 We further designed a method to generate a uniform {\em a priori} stellar population distribution.
 The same method allows us to transform the complex hyper-tetrahedron boundary constraints into simple bounds constraints.
 This transformation simplifies considerably the search, when needed, for a minimum in the stellar population space.

\section*{Acknowledgments}
 We wish to thank Yvon Biraud, Catherine Boisson, Silvano Bonazzola, Monique Joly, Jean-Alain Marck and Dr. Yuen Keong Ng whose comments about this paper have been so helpful.
 Special thanks go to Dr. Richard Barrett who suggested the concept of information developed here.

\appendix
\section{Uniform mapping of a simplex}
\label{ap.1}
 In this appendix we show how to transform an hyper-cube into an hyper-tetrahedron.
 This operation is continuous and transforms a uniform density on the hyper-cube into a density equally uniform on the hyper-tetrahedron.

 The transformation $\psi$ allows a simplification of the constraints $\bm{k}\ge \bm{0}$ subject to $\sum_i k_i=1$ ($\bm{k}\in{\cal S}$) through the mapping $\bm{k}=\psi(\bm{u})$ of the variables $u_i$ subject to simple bounds constraints of the type: $0\le u_i \le 1$.
 At the same time the uniform coverage of the hyper-tetrahedron ${\cal S}$ permits to evaluate the information integral~(\ref{eq.info}).

 We also want the change of variables $\psi$ to be continuous and twice derivable, so that $\psi$ can be used in an optimization program where the variables are subject to the constraints $\bm{k}\in{\cal S}$ as needed in \S~\ref{ch.1.2}.
 
 One can think of several ways to map $\bm{k}=\psi(\bm{u})$, but if the transformation is not chosen carefully there is a risk that $\psi$ selects preferentially certain zones of the $\bm{k}$ parameters space.
 In order to avoid this possible bias, we would like to find a change of variable ensuring that a uniform coverage of the $\bm{u}$ space (the cube) transforms to a uniform coverage of the $\bm{k}$ space (the tetrahedron).
 We found two methods satisfying these requirements.

\subsection{Method 1.}
 Let us consider the $\ns$ independent random variables $K_i$, following all the same exponential distribution.
 This distribution is, for $k\geq 0$ a probability density function: 
$f(k)=\exp(-k)$.
 The joint distribution of the multiplet $(K_1,\ldots,K_{\ns})$ has the density $f_{\ns}(k_1,\ldots,k_{\ns})=\prod_{i=1}^{\ns}f(k_i)=\exp(-\sumi k_i)$ and is constant on $\sumi k_i=a$.
 The conditional density on $\sumi k_i=a$ is therefore uniform for any $a>0$, in particular for $a=1$.

 Now the random variable $U_i=\int_0^{K_i}f(k)\,dk=1-\exp(-K_i)$ is uniformly distributed on $0\leq u_i<1$ so, $K_i=-\ln(1-U_i)$ or equivalently $K_i=-\ln(U_i)$ is exponentially distributed.
  In conclusion, the mapping $\psi$ defined by:
\beq
 k_i = -\ln(u_i)\,,\quad i=1,\ldots,\ns\,,\quad 0<u_i\leq1\,,
\eeq
 ensures: (i) $k_i \geq 0$; (ii) a uniform coverage of the $k_i$ domain; and 
(iii) that $u_i$ is only subject to simple bounds constraints $0\leq u_i<1\,, 
\;i=1,\ldots,\ns$.

 The only problem with this method is that we use $\ns$ parameters 
while only $\ns-1$ are necessary.
 The search algorithm may loose time exploring those lines in $\bm{u}$ space 
where $\Wsynj$ does not change because the $k_i$ differ only by a scaling 
factor.
 We thus designed `method 2' in order to solve this problem.

\subsection{Method 2.}
 Here we want to map the $k_i$ using only $\ns-1$ parameters of type $u_i$.
 Again the $K_i$ are exponentially distributed as in `method 1'; this ensures, 
as shown above, that the density is uniform on $\sumi k_i=1$.

 We now introduce the new variables $Q_i$. For the sake of argument we limit ourself to a $\ns=4$ example, but it should be clear that the method works for any $\ns>0$.
 We define:
\[
\begin{array}{r@{=\,}ll}
 Q_1 & K_4 + K_3 + K_2 + K_1\,,&\quad 0\leq Q_1 < \infty\,;\\
Q_2Q_1 & K_4 + K_3 + K_2\,,&\quad 0\leq Q_2\leq1\,;\\
Q_3Q_2Q_1 & K_4 + K_3\,,&\quad 0\leq Q_3\leq1\,;\\
Q_4Q_3Q_2Q_1 & K_4\,,&\quad 0\leq Q_4\leq1\,.
\end{array}
\]
 The Jacobian of this change of variables, 
$J_4=\dfrac{\partial(k_1,\ldots,k_4)}{\partial(q_1,\ldots,q_4)}$, is given by: 
$J_4=q_1^3q_2^2q_3$. The four values $(Q_1,Q_2,Q_3,Q_4)$ have the density:
\beq
 f(q_1,q_2,q_3,q_4)=e^{-q_1}q_1^3q_2^2q_3\,.
\eeq
 This demonstrates that the $Q_i$ are {\em independent} and have densities:
\beq
\left.
\begin{array}{l@{\;=\;}ll}
f_1(q_1)&\dfrac{1}{3!}q_1^3e^{-q_1}\,, &0\leq q_1<\infty\,;\\[1ex]
f_2(q_2)&3q_2^2\,, &0\leq q_2 <1\,;\\[1ex]
f_3(q_3)&2q_3\,,   &0\leq q_3 <1\,;\\[1ex]
f_4(q_4)&1\,, &0\leq q_4 <1\,.
\end{array}
\right\}
\eeq
 The variable $Q_1$ follows a gamma distribution, (as required for a sum of independent exponential random variables) and the other variables follow a power distribution of decreasing index, the last one being uniform.

\begin{figure}
\centering
\psfig{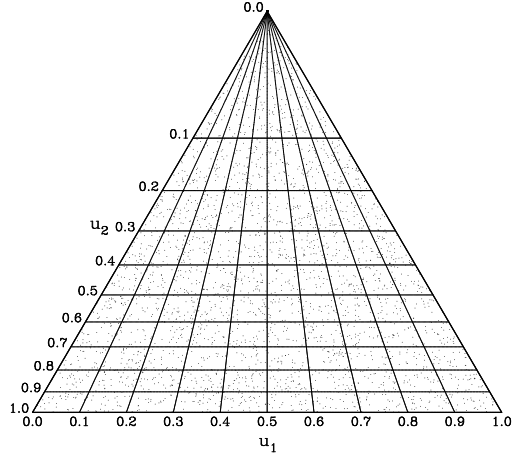}
\caption{Uniform filling of an equilateral triangle (i.e. an hyper-tetrahedron 
in a 3D space) by means of `method 2'.
The points are obtained by using two independent random variables $U_1$ and 
$U_2$ uniformly distributed over $[0,1]$.
According to Eqs.~(\ref{eq.a7}) and~(\ref{eq.a8}) the change of variables is: 
$k_1=1-\sqrt{u_2}$, $k_2=(1-u_1)\sqrt{u_2}$ and 
$k_3=u_1\sqrt{u_2}$.}
\label{fig.a1}
\end{figure}

 Now if $Q_1$ is given, say $Q_1\equiv \sumi K_i=1$, the remaining $Q_2$, $Q_3$
and $Q_4$ cover uniformly (because the $K_i$ are exponential) the 
hyper-tetrahedron defined by (\ref{eq.setS}).
 We have:
\beq
\left.
\begin{array}{l@{\;=\;}l}
K_4&Q_4Q_3Q_2\,,\\[1ex]
K_3&(1-Q_4)Q_3Q_2\,,\\[1ex]
K_2&(1-Q_3)Q_2\,,\\[1ex]
K_1&1-Q_2\,.
\end{array}
\right\}
\eeq
 Finally if we define:
\beq
 U_n =\hspace{-2ex} \int\limits_0^{Q_{\ns-n+1}} \hspace*{-1em}n q^{n-1}dq=Q_{\ns-n+1}^n\,,
\quad n=1,\ldots,\ns-1\,,
\eeq
the variables $U_n$ are uniform and independent.
 We then have a power distribution of index $n$: $Q_{\ns-n+1}=U_n^{\frac{1}{n}}$ (for $n=1,\ldots,\ns-1$) when $U_n$ is uniformly distributed in $[0,1]$.
 Therefore the change of variables $\psi$:
\beq
\left.
\begin{array}{l@{\;=\;}l}
k_4&u_1 u_2^{\frac{1}{2}}u_3^{\frac{1}{3}}\,,\\
k_3&(1-u_1)u_2^{\frac{1}{2}}u_3^{\frac{1}{3}}\,,\\
k_2&(1-u_2^{\frac{1}{2}})u_3^{\frac{1}{3}}\,,\\
k_1&1-u_3^{\frac{1}{3}}\,,
\end{array}
\right\}
\eeq
maps the $\bm{k}$ space uniformly if $u_1,u_2$ and $u_3$ are drawn from three independent random variables $U_1,U_2,U_3$ uniformly distributed in $[0,1]$.
 Now for any $\ns > 1$, $\psi$ is defined by:
\beq\label{eq.a7}
k_1 = 1-u_{\ns-1}^{1/(\ns-1)}\,,
\quad
k_{\ns}=\prod_{i=1}^{\ns-1}u_i^{\frac{1}{i}}\,,
\eeq
and for $n=2,\ldots,\ns-1$:
\beq\label{eq.a8}
 k_n=[1-u_{\ns-n}^{{1}/{(\ns-n)}}]
 \prod_{i=\ns-n+1}^{\ns-1}\hspace*{-1em}u_i^{\frac{1}{i}}\,,
\eeq
 where all $u_i$ are subject to the constraint: $0\leq u_i <1$.
 We present in Fig.~\ref{fig.a1} an example of such a mapping for 
\mbox{$\ns=3$}.

\section{Computations needed for the error analysis.}\label{ap.2}

 We would like to establish a relationship between a small variation $d\bm{W}$ around $\bm{\Wobs}$ and the resulting variation $d\bm{k}$ around the solution $\bm{k}$.
 The variation $d\bm{k}$ is constrained to $\sumi dk_i=0$ by $\sumi k_i=1$.

\subsection{The galaxy can be synthesized.}\label{ch.B1}

 $\bm{\Wobs}$ belongs to the synthetic domain ($\bm{\Wobs}\in {\cal W}$) if the galaxy can be synthesized and the many solutions are contained within the convex hull of a set of extreme solutions (see \paperII).
 An extreme solution has by definition at least $\ns-(\nl+1)$ components $k_i$ equal to zero, what remains is the evaluation of $\nl+1$ components of $\bm{k}$.

\subsubsection{Error analysis of the extreme solutions.}\label{ap.211}

 We discard the stars with zero contribution and set $\ns=\nss$ (recall that $\nss=\nl+1$).
 This allows us to keep the same notation. Note that the new matrix $\bM{A}$ is formed of the $\nl+1$ columns of the original matrix in system~(\ref{eq.Bk}) corresponding to the stars which have not been set to zero.
 In the same way $\varphi$ stands here for the original $\varphi$ restricted to these stars.

 By adding the line $\bm{b}^T$ to $\bM{A}$ we form a $\nl+1$ by $\nl+1$ matrix $\bM{B}$ and $\bM{B}^{-1}$ exists, because the problem has an (extreme) solution.
 We have:
\beq\label{eq.B1}
\bm{k} = \bM{B}^{-1} \parl\begin{array}{c}\bm{0}\\1\end{array}\parr\,,
\quad
\bM{B} =\Bigl(\!\begin{array}{l}\bM{A}\\ \bm{b}^T\end{array}\!\Bigr)\,.
\eeq
 From here we have $d\bM{B}\bm{k}+\bM{B}d\bm{k}=\bm{0}$, the matrix $d\bM{B}$ is $d\bM{A}$ supplemented with a line of zeros.
 We have $dA_{ji}=dW_j I_{ji}$ and $[d\bM{A}\bm{k}]_j=dW_j\sumi I_{ji}k_i=dW_j\Isynj$.
Let us introduce the diagonal matrix $\bM{J}$, where $J_{jj}=\Isynj$ for $j=1,\ldots,\nl$ and the last diagonal element is $J_{\nl+1,\nl+1}=1$.
 We then have:
\beq\label{eq.B2}
 \bM{B}d\bm{k} = -d\bM{B}\bm{k}
 =-\bM{J}\parl\begin{array}{c}d\bm{W}\\0\end{array}\parr\,.
\eeq
As none of the continua $I_{ji}$ are null, $\bM{J}^{-1}$ exists and we define $\bM{W}=\bM{J}^{-1}\bM{B}$.
 The matrix $\bM{W}$ has a very simple form. For example: if we have $\nl=2$, the number of stars of an extreme ray is $\ns=3$ and $\bM{W}$ equals:
\[
 \left(\!\!
 \begin{array}{ccc}
  {\scriptscriptstyle (W_{\obs 1}-W_{11})\frac{I_{11}}{I_{\syn 1}}} \!&
  {\scriptscriptstyle (W_{\obs 1}-W_{12})\frac{I_{12}}{I_{\syn 1}}} \!&
  {\scriptscriptstyle (W_{\obs 1}-W_{13})\frac{I_{13}}{I_{\syn 1}}} \\[1ex]
  {\scriptscriptstyle (W_{\obs 2}-W_{21})\frac{I_{21}}{I_{\syn 2}}} \!&
  {\scriptscriptstyle (W_{\obs 2}-W_{22})\frac{I_{22}}{I_{\syn 2}}} \!&
  {\scriptscriptstyle (W_{\obs 2}-W_{23})\frac{I_{23}}{I_{\syn 2}}} \\
  1 & 1 & 1
 \end{array}
       \!\!\right).
\]
 If we define $\bM{K}=\bM{W}^{-1}=\bM{B}^{-1}\bM{J}$, equation~(\ref{eq.B2}) can be written as
\beq\label{eq.B21}
 \bM{W}d\bm{k} = -\parl\begin{array}{c}d\bm{W}\\0\end{array}\parr
\quad\mbox{or}\quad
 d\bm{k} = -\bM{K}\parl\begin{array}{c}d\bm{W}\\0\end{array}\parr\,.
\eeq
 We can now compute the \vaco\ matrix of the $\bm{k}$'s.
 Let $\bM{V}_{k}$ be that matrix. In the tangent approximation we have $\bM{V}_{k}=\e{d\bm{k}d\bm{k}^T}$:
\[
\begin{array}{l@{\;=\;}l}
\bM{V}_{k}&\e{\bM{K}\parl\begin{array}{c}d\bm{W}\\0\end{array}\parr
(d\bm{W}^T\, 0)\bM{K}^T}\,,\\[1ex]
 &\bM{K}\parl
 \begin{array}{cc} \e{d\bm{W}d\bm{W}^T}&\bm{0}\\\bm{0}^T&0\end{array}
 \parr\bM{K}^T\,.
\end{array}
\]
 If $\bM{V}$ designates the \vaco\ matrix of the observation we finally have:
\beq\label{eq.B3}
 \bM{V}_k = \bM{K}\parl
 \begin{array}{lc}\bM{V}&\bm{0}\\\bm{0}^T&0\end{array}
 \parr\bM{K}^T\,.
\eeq

 As a bonus, we note that it is possible to express the solution $\bm{k}$ in terms of $\bM{K}$.
 We have $\bM{B}=\bM{J}\bM{W}$, $\bM{B}^{-1}=\bM{K}\bM{J}^{-1}$ and the solution~(\ref{eq.B1}) can be expressed by:
\[
 \bm{k} = 
\bM{K}\bM{J}^{-1}\parl\begin{array}{c}\bm{0}\\1\end{array}\parr =
\bM{K}           \parl\begin{array}{c}\bm{0}\\1\end{array}\parr\,.
\]

\subsubsection{Acceptance region of the extreme solutions}\label{ch.B12}

 At the level $\gamma$, all points around the solution $\bm{k}$ are indistinguishable and give an image close to the observation \bm{\Wobs}.
 The acceptable solutions $\bm{k}'$ are in a set ${\cal K}_\gamma$ called the `acceptance region'.
 We define: ${\cal K}_\gamma = \varphi^{-1}({\cal E}_\gamma)$.

 In the tangent approximation around $\bm{k}$, we have:
\[
{\cal K}_\gamma = \set{\bm{k}'}{(\bm{k}'-\bm{k})^T\bM{P}(\bm{k}'-\bm{k})\le F^{-1}_{\chi^2}(\gamma)}\,,
\]
 where, as in  Eq.~(\ref{eq.setE}), $F_{\chi^2}$ is the repartition function of a $\chi^2$ variate possessing $\nl$ degrees of freedom.
 That is also the equation of an ellipsoid around $\bm{k}$.
 The `weight' matrix $\bM{P}$ is usually equal to the inverse of the \vaco\ matrix $\bM{V}_k$.
 The problem is that $\bM{V}_k$, as it is clear from~(\ref{eq.B3}), does not possess an inverse. 
 Back to the definition of $\varphi^{-1}({\cal E}_\gamma)$ we have:
\[
\begin{array}{l@{\;=\;}l}
{\cal K}_\gamma & \set{\bm{k}'}
{[\varphi(\bm{k}')-\varphi(\bm{k})]^T
\bM{V}^{-1}
[\varphi(\bm{k}')-\varphi(\bm{k})]
\le F^{-1}_{\chi^2}(\gamma)}\,,\\[1ex]
& \set{\bm{k}'}{d\bm{W}^T\bM{V}^{-1}d\bm{W}\le F^{-1}_{\chi^2}(\gamma)}\,,
\end{array}
\]
where we have set $d\bm{W}=\varphi(\bm{k}')-\varphi(\bm{k})$.
 The above expression is equivalent with:
\[
\begin{array}{l@{\;=\;}l}
{\cal K}_\gamma & \set{\bm{k}'}
{(d\bm{W}^T 0)
\parl\begin{array}{lc}\bM{V}^{-1}&\hspace{-2ex}\bm{0}\\ 
                      \bm{0}^T&\hspace{-2ex} 0\end{array}\parr
\parl\begin{array}{c}d\bm{W}\\0\end{array}\parr
\le F^{-1}_{\chi^2}(\gamma)}\,.
\end{array}
\]
 The choice of the extra elements in the central matrix is arbitrary.
 We chose zero for simplicity.
 We can now use Eq.~(\ref{eq.B21}) and write:
\[
{\cal K}_\gamma = \set{\bm{k}'}
{(\bm{k}'-\bm{k})^T\bM{W}^T
\parl\begin{array}{lc}\bM{V}^{-1}&\hspace{-2ex}\bm{0}\\
                      \bm{0}^T&\hspace{-2ex} 0\end{array}\parr
\bM{W}(\bm{k}'-\bm{k})
\le F^{-1}_{\chi^2}(\gamma)}\,.
\]
 Therefore:
\[
\bM{P} =
\bM{W}^T
\parl\begin{array}{lc}\bM{V}^{-1}&\hspace{-2ex}\bm{0}\\
                      \bm{0}^T&\hspace{-2ex} 0\end{array}\parr
\bM{W}\,.
\]

\subsubsection{Contributions to errors coming from stellar library}\label{ap.213}

 Let's call $\bM{J}_{Wi}$ and $\bM{J}_{Ii}$ the diagonal matrices such that for $j=1,\ldots,\nl$ we have $[\bM{J}_{Wi}]_{jj}=I_{ji}k_i$ and $[\bM{J}_{Ii}]_{jj}=(\Wsynj-W_{ji})k_i$. In addition we have for $j=\nl+1$ that $[\bM{J}_{Wi}]_{jj}=[\bM{J}_{Ii}]_{jj}=1$.
 Define further $\bM{K}_{Wi}=\bM{B}^{-1}\bM{J}_{Wi}$ and $\bM{K}_{Ii}=\bM{B}^{-1}\bM{J}_{Ii}$
 Following the same reasoning as in section~\ref{ap.211} we find that the contribution of the stellar library to the total \vaco\ matrix of a solution is:
\[
\bM{V}_{k(\star)} =
 \sum_{i=1}^{\nl+1}\{
 \bM{K}_{Wi}
 \parl
 \begin{array}{lc}
 \bM{V}_{Wi}&\hspace{-2ex}\bm{0}\\
 \bm{0}^T   &\hspace{-2ex}0
 \end{array}
 \parr
 \bM{K}_{Wi}^T+
 \bM{K}_{Ii}
 \parl
 \begin{array}{lc}
 \bM{V}_{Ii}&\hspace{-2ex}\bm{0}\\
 \bm{0}^T   &\hspace{-2ex}0
 \end{array}
 \parr
 \bM{K}_{Ii}^T\}\,.
\]
 Where $\bM{V}_{Wi}$ and $\bM{V}_{Ii}$ are the \vaco\ matrices of the $W_{ji}$ and $I_{ji}$ of star $i$.
 A comparison of $\bM{V}_{k(\star)}$ and $\bM{V}_k$ is able to tell if the errors introduced by the stellar library are negligible relative to the errors on the observed galaxy.

\subsection{The galaxy cannot be synthesized}

 An approximate solution $\bm{\Wsyn}$ is found if the galaxy cannot be synthesized. A manifold of solutions is defined by a number of stars $\nss$ less or equal to the number of \eqws.
 Here we distinguish $\nss$ from the $\ns$ number of stars in the data base.
 This manifold is called the synthetic `surface' in \paperI.

\subsubsection{Projector on the tangent plane.}
 If the observational errors $d\bm{\Wobs}$ are small around $\bm{\Wobs}$, we can expect small deviations $d\bm{\Wsyn}$ around $\bm{\Wsyn}$.
 In this context, we can approximate the synthetic surface by a plane tangent to this surface at $\bm{\Wsyn}$.
 The results obtained in \ref{ch.B1} can be applied to $\bm{\Wsyn}$ since, by definition, $\bm{\Wsyn}$ belongs to the synthetic domain.
 In this context let's $\dot{\bM{A}}$ and $\dotB$ stand for the matrices $\bM{A}$ and $\bM{B}$, where $\bm{\Wobs}$ have been replaced by $\bm{\Wsyn}$.
 By (\ref{eq.B2}) $\bm{\Wsyn} + d\bm{\Wsyn}$ belongs to the tangent plane if the $d\bm{k}$ are such that:
\beq\label{eq.B05}
\parl\begin{array}{c}d\bm{\Wsyn}\\0\end{array}\parr=
 -\bM{J}^{-1}\dotB d\bm{k}
\quad{\mbox{or}}\quad
d\bm{\Wsyn}=
 -\bM{J}_A^{-1}\dot{\bM{A}}d\bm{k}
\,,
\eeq
with $\bM{J}_A=\diag(I_{\syn 1},\ldots,I_{\syn\nl})$.
 This means that $d\bm{\Wsyn}$ belongs to the linear hull of the columns of $\bM{J}_A^{-1}\bM{A}$ {\em provided} that $\sum_i dk_i=0$.
 We maintain that $d\bm{\Wsyn}$ belongs to this linear hull {\em even} if $\sum_idk_i\not=0$.

 In fact $\sum_idk_i=0$ is the equation of an hyperplane $\Sigma_0$ orthogonal to the subspace of dimension one generated by e.g. the vector $(1,\ldots,1)$.
 The hyperplane $\Sigma_0$ is itself a subspace since it includes the origin, therefore any vector $d\bm{k}$ is the sum of a vector $d\bm{k}_0\in\Sigma_0$ plus a vector $\bm{v}\not\in\Sigma_0$.
 By construction the solution $\bm{k}\not\in\Sigma_0$ since $\sum_ik_i=1$, then $d\bm{k}$ can be written $d\bm{k}=d\bm{k}_0+d\alpha\bm{k}$, where $d\alpha$ is a scalar.
 Now we have:
\[
\begin{array}{r@{\;=\;}l}
\bM{J}^{-1}\dotB d\bm{k} &
\bM{J}^{-1}\dotB(d\bm{k}_0+d\alpha\bm{k})
\,,\\[0.8ex]
 & 
\bM{J}^{-1}\dotB d\bm{k}_0+d\alpha\bM{J}^{-1}\dotB\bm{k}
\,,\\[0.8ex]
 &
\parl\begin{array}{c}d\bm{\Wsyn}\\0\end{array}\parr
+
\parl\begin{array}{c}0\\d\alpha\end{array}\parr
\,.
\end{array}
\]
 Therefore $d\bm{\Wsyn}$ belongs to the linear hull of the columns of $\bM{G}=\bM{J}_A^{-1}\dot{\bM{A}}$ i.e. it belongs to the {\em  range} of $\bM{G}$.

 The range of $\bM{G}$ is obtained by the rank deficiency QR algorithm (see Golub and Van~Loan 1996 Chapter~5 \S5.4.1).   
 According to this algorithm it is possible to write, after a permutation $\bm{\Pi}$, $\bM{G}\bm{\Pi}=\bM{Q}\bM{R}$.
 The array $\bM{Q}$ is a $\nl$ by $\nl$ matrix.
 The $\nss-1$ first columns of $\bM{Q}$ span the range of $\bM{G}$ (i.e. they are an orthonormal basis of the tangent plane) and the remaining columns span the subspace orthonormal to it.
 If, following this partition, we write $\bM{Q}=(\bM{p}|\bM{q})$, one can construct the orthogonal projector $\bM{H}$ on the tangent plane, we have:
\[
\bM{H} = \bM{p}\bM{p}^T\,.
\]
 The projector $\bM{H}$ is a $\nl$ by $\nl$ matrix of rank $\nss-1$.

\subsubsection{Least square solution of the linearized problem}
  One obtains $\bm{W}'_{\syn}=\bm{\Wsyn}+d\bm{\Wsyn}$ from $\bm{W}'_{\obs}=\bm{\Wobs}+d\bm{\Wobs}$ by minimizing a `distance' between these $\bm{W}'_{\syn}$ and $\bm{W}'_{\obs}$. 
 This distance is usually defined as a positive definite quadratic form, therefore the correction $d\bm{\Wsyn}$ is implicitly defined by:
\begin{eqnarray}
\lefteqn{
(\bm{W}'_{\obs}-\bm{W}'_{\syn})^T\bm{\Sigma}^{-1}
(\bm{W}'_{\obs}-\bm{W}'_{\syn}) =
}
 \hspace{4.5em}\nonumber \\
& & \min_{\bm{W}'} (\bm{W}'_{\obs}-\bm{W}')^T\bm{\Sigma}^{-1}
                  (\bm{W}'_{\obs}-\bm{W}')\,.\label{eq.B5}
\end{eqnarray}
 Via a change of variables, one can describe the `weight' matrix $\bm{\Sigma}^{-1}$ by the identity matrix (see Sect. \ref{ch.B26}). 
 We hereafter consider that $\bm{\Sigma}^{-1}=\bM{I}$, then the synthesis is obtained by the orthogonal projection of $\bm{\Wobs}$ on the synthetic surface i.e. using $\bM{H}$. 
 Consequently one obtains $\bm{W}'_{\syn}$ by adding $d\bm{\Wsyn}=\bM{H}(\bm{W}'_{\obs}-\bm{W}_{\syn})$ to $\bm{\Wsyn}$.
 Taking into account that $\bM{H}(\bm{\Wobs}-\bm{\Wsyn})=0$ we obtain:
\beq\label{eq.B6}
 d\bm{\Wsyn}=\bM{H}d\bm{\Wobs}\,.
\eeq

\subsubsection{\Vaco\ matrix of \bm{k}}
 The second step is to find a relation between $d\bm{\Wsyn}$ and $d\bm{k}$ where $d\bm{k}$ is restricted to the $\nss$ stars which contribute to the synthesis.
 Since, by definition $\bm{\Wsyn}\in{\cal W}$, the overdetermined system $\dotB\bm{k}=(\bm{0}\,1)^T$ possesses an exact solution in the least-square sense.
 Therefore $\bm{k}$ is also the solution of the invertible square system $\dotB^T\dotB\bm{k}=\dotB^T(\bm{0}\,1)^T$.
 In this linear analysis, the perturbed $\bm{W}'_{\syn}$ is close to the synthetic surface and the normalization constraint $\bm{b}^T\bm{k}=1$ is approximately satisfied.
 It follows that $\dotB^Td\dotB\bm{k}+\dotB^T\dotB d\bm{k}=0$. Again $d\dotB\bm{k}$ is given by Eq.~(\ref{eq.B2}), where $\bM{J}$ is the diagonal matrix of the synthetic continua $\bM{J}=\diag(I_{\syn 1},\ldots,I_{\syn\nl},1)$.
 This leads to:
\beq\label{eq.B7}
d\bm{k} = -(\dotB^T\dotB)^{-1}\dotB^T\bM{J}
 \parl\begin{array}{c}d\bm{\Wsyn}\\0\end{array}\parr\,.
\eeq
 Note that $\dotB$ being of full column rank, the expression $(\dotB^T\dotB)^{-1}\dotB^T$ is equal to $\dotB^{(-1)}$ the Moore-Penrose pseudo-inverse of $\dotB$ (for a definition see e.g. Harville 1997 Chap.~20.1).
 Let's define here $\bM{K} = (\dotB^T\dotB)^{-1}\dotB^T\bM{J}$ and recalling that $d\bm{\Wsyn}=\bM{H}\,d\bm{\Wobs}$ we obtain the \vaco\ matrix of the synthesis:
\beq\label{eq.B8}
\bM{V}_k = \bM{K}
\parl
 \begin{array}{cc}\bM{H}\bM{V}\bM{H}^T&\hspace{-2ex}\bm{0}\\
 \bm{0}^T&\hspace{-2ex}0\end{array}
 \parr
\bM{K}^T\,,
\eeq
where $\bM{V}=\e{d\bm{\Wobs}d\bm{W}^T_{\obs}}$ is the \vaco\ matrix of the observations. 

\subsubsection{Acceptance region of the least-square solutions}
 An alternative solution $\bm{k}'$ is considered acceptable, at the level $\gamma$, if there exists a $\bm{W}'_{\obs}$ within the `error' region ${\cal E}_{\gamma}$ around $\bm{\Wobs}$ such that $\bM{H}\bm{W}'_{\obs}=\bm{W}'_{\syn}=\varphi(\bm{k}')$.
 According to this definition, $\bm{k}'$ is acceptable if it belongs to the following set:
\beq
\left.
\begin{array}{l}
{\cal K}_\gamma =
 \set{\bm{k}'}{d\bm{W}^T_{\obs}\bM{V}^{-1}d\bm{\Wobs}\le F^{-1}_{\chi^2}(\gamma)}\,, \\[1ex]
 \bM{H}d\bm{\Wobs}=d\bm{\Wsyn}=\varphi(\bm{k}')-\varphi(\bm{k})\,,
\end{array}\right\}
\eeq
where $\bM{V}$ is the \vaco\ matrix of the observational errors.

 Among all the solutions of $\bM{H}d\bm{\Wobs}=d\bm{\Wsyn}$ we choose the one which is `closest' to $\bm{\Wobs}$ in the $\bM{V}^{-1}$-norm.
 This ensures that the solution is within ${\cal E}_\gamma$.
 The minimum of $d\bm{W}^T_{\obs}\bM{V}^{-1}d\bm{\Wobs}$, which satisfies the constraint $\bM{H}d\bm{\Wobs}=d\bm{\Wsyn}$ is given by:
\[
\begin{array}{r@{\;=\;}l}
d\bm{\Wobs} &\bM{M}\,d\bm{\Wsyn}\,,\\[0.8ex]
\bM{M} & \bM{pp}^T+\bM{qq}^T\bM{V}\bM{p}(\bM{p}^T\bM{V}\bM{p})^{-1}\bM{p}^T\,.
\end{array}
\]
 The matrix $\bM{M}$ is an oblique projector on the regression plane defined by the constraint.
We indeed have:
\beq\label{eq.B11}
 \parl\begin{array}{c}d\bm{\Wobs}\\0\end{array}\parr =
 \parl\begin{array}{c}\bM{M}\,d\bm{\Wsyn}\\0\end{array}\parr 
\,.
\eeq
 
 One cannot invert Eq.~(\ref{eq.B7}) in order to introduce $d\bm{k}$, but we can use Eq.~(\ref{eq.B05}) since ${\cal K}_\gamma$ is in the set where $\sum_idk_i=0$:
\beq\label{eq.B10}
 \parl\begin{array}{c}d\bm{\Wsyn}\\0\end{array}\parr =
-\bM{J}^{-1}\dotB\,d\bm{k}\,.
\eeq
 Combining Eqs.~(\ref{eq.B10}) and~(\ref{eq.B11}) and proceeding like in Sect. \S~(\ref{ch.B12}), we find:
\beq
{\cal K}_\gamma =
\set{\bm{k}'}{(\bm{k}'-\bm{k})^T\bM{P}(\bm{k}'-\bm{k})
\le F^{-1}_{\chi^2}(\gamma)}\,,
\eeq
\beq\label{eq.B14}
 \bM{P} = \bM{W}^{T}
\parl\begin{array}{cc}
 \bM{M}^T\bM{V}^{-1}\bM{M} & \hspace{-1.0ex}\bm{0}\\
 \bm{0}^T&\hspace{-1.0ex} 0
\end{array}
\parr
\bM{W}\,,
\quad
\bM{W}=\bM{J}^{-1}\dotB\,.
\eeq

\subsubsection{Weighted least square}\label{ch.B26}
 We consider now the case where the distance between two points of the $W$-space is not the Euclidean distance but an elliptical distance.
 This elliptical distance is subordinated to a symmetric definite positive matrix $\bm{\Sigma}$.
 We have $d_\Sigma(\bm{W}_1,\bm{W}_2)=(\bm{W}_1-\bm{W}_2)^T\bm{\Sigma}^{-1}(\bm{W}_1-\bm{W}_2)$.
 Usually one chooses $\bm{\Sigma}$ equal to $\bM{V}$ the \vaco\ matrix of the data, but it may not be always the case.

 A substitution of variables transforms the elliptical distance into a spherical (i.e. Euclidean) one.
 We want in other words that $d(\bm{W}_1,\bm{W}_2)=d_\Sigma(\bm{W}_{10},\bm{W}_{20}), $where $d$ is the Euclidean distance. If $\bm{W}_0$ stands for the {\em original} variables and $\bm{W}$ for the {\em new} variables, we define the matrix $\bM{N}$ by: $\bm{W}_0=\bM{N}\bm{W}$. It now follows that
\[
d(\bm{W}_1,\bm{W}_2)=
(\bm{W}_1-\bm{W}_2)^T\bM{N}^T\bm{\Sigma}^{-1}\bM{N}(\bm{W}_1-\bm{W}_2)\,.
\]
 This imposes the condition $\bM{N}^T\bm{\Sigma}^{-1}\bM{N}=\bM{I}$, which is satisfied if we choose $\bM{N}$ as the matrix appearing in the Choleski factorization of $\bm{\Sigma}$ (see Golub \&\ Van Loan 1996, section 4.2.1).
 We have $\bm{\Sigma}=\bM{N}\bM{N}^T$, where $\bM{N}$ is lower triangular. It is straightforward to verify that $\bM{N}$ satisfies the condition.  

 Finally we need to compute the \vaco\ matrix $\bM{V}$ appearing in Eqs.~(\ref{eq.B8}) and~(\ref{eq.B14}) from the \vaco\ matrix $\bM{V}_0$ of the original variables.
 We have $\bM{V}=\e{d\bm{W}d\bm{W}^T}=\bM{N}^{-1}\e{d\bm{W}_0\bm{W}_0^T}\bM{N}^{-1T}$, therefore:
\beq
\bM{V} = \bM{N}^{-1}\bM{V}_0\bM{N}^{-1T}\,.
\eeq

\begin{table*}
\caption
{The Bica solution and our 16 extreme solutions of galaxy NGC\,3521 synthesis using as data base the 8 clusters (2--7,9,15) retain by E.~Bica in his synthesis.
 Under column $k_i$  is the contribution to the galactic luminosity of cluster Nr.~$i$; $\sigma\bm{k}_s, (s=1,2,...16)$ is the standard deviation corresponding to the extreme solution $\bm{k}_s$ supposing only errors on galactic data and $\sigma'\bm{k}_s$ is the standard deviation corresponding to the same extreme solution taking into account errors on data base.
 Finally, `surface' is the value of the ellipsoid's surface for each extreme solution.
The solutions are sorted according to a decreasing order of merit (increasing order of surface).
}
\label{tab.2}
\input{tab2}

\end{table*}
\end{document}

%% file: tab2.tex
\begin{tabular}{rrrrrrrrrc}
\hline
\multicolumn{1}{r}{ } &
\multicolumn{1}{c}{$k_2$} &
\multicolumn{1}{c}{$k_3$} &
\multicolumn{1}{c}{$k_4$} &
\multicolumn{1}{c}{$k_5$} &
\multicolumn{1}{c}{$k_6$} &
\multicolumn{1}{c}{$k_7$} &
\multicolumn{1}{c}{$k_9$} &
\multicolumn{1}{c}{$k_{15}$} &
surface \\
\hline
$\bm{k}_{\mbox{Bica}}$ & 
 0.510 & 0.090 & 0.080 & 0.070 & 0.060 & 0.050 & 0.080 & 0.060 \\[0.8ex]
$\bm{k}_1$ &
 0.561 & 0.000 & 0.000 & 0.000 & 0.000 & 0.091 & 0.000 & 0.348 & 6.462$\,10^{-5}$\\
$\sigma\bm{k}_1$ &
$\pm$0.334 & & & & & $\pm$0.067 & & $\pm$0.407 \\
$\sigma'\bm{k}_1$ &
$\pm$0.353 & & & & & $\pm$0.069 & & $\pm$0.418 \\[0.8ex]

$\bm{k}_2$ &
 0.000 & 0.875 & 0.000 & 0.000 & 0.000 & 0.075 & 0.000 & 0.050 & 7.779$\,10^{-5}$\\
$\sigma\bm{k}_2$ &
 &$\pm$0.454 & & & & $\pm$0.048 & & $\pm$0.498 \\
$\sigma'\bm{k}_2$ &
 &$\pm$0.474 & & & & $\pm$0.050 & & $\pm$0.520 \\[0.8ex]

$\bm{k}_3$ &
 0.390 & 0.000 & 0.601 & 0.000 & 0.000 & 0.009 & 0.000 & 0.000 & 8.759$\,10^{-5}$\\
$\sigma\bm{k}_3$ &
$\pm$0.451 & & $\pm$0.581 & & & $\pm$0.132 & & \\
$\sigma'\bm{k}_3$ &
$\pm$0.463 & & $\pm$0.598 & & & $\pm$0.136 & & \\[0.8ex]

$\bm{k}_{4}$ &
 0.570 & 0.000 & 0.000 & 0.000 & 0.146 & 0.000 & 0.000 & 0.284 & 1.456$\,10^{-4}$\\
$\sigma\bm{k}_{4}$ &
$\pm$0.339 & & & &$\pm$0.104 & & &$\pm$0.438 \\
$\sigma'\bm{k}_{4}$ &
$\pm$0.347 & & & &$\pm$0.106 & & &$\pm$0.448 \\[0.8ex]

$\bm{k}_5$ &
 0.000 & 0.868 & 0.000 & 0.086 & 0.000 & 0.046 & 0.000 & 0.000 & 2.085$\,10^{-4}$\\
$\sigma\bm{k}_5$ &
 &$\pm$0.515 & &$\pm$0.842 & & $\pm$0.328 & &  \\
$\sigma'\bm{k}_5$ &
 &$\pm$0.538 & &$\pm$0.880 & & $\pm$0.343 & & \\[0.8ex]

$\bm{k}_6$ &
 0.267 & 0.000 & 0.000 & 0.000 & 0.000 & 0.165 & 0.568 & 0.000 & 2.121$\,10^{-4}$\\
$\sigma\bm{k}_6$ &
$\pm$0.724 & & & & & $\pm$0.032 & $\pm$0.700 &  \\
$\sigma'\bm{k}_6$ &
$\pm$0.748 & & & & & $\pm$0.033 & $\pm$0.723 &  \\[0.8ex]

$\bm{k}_7$ &
 0.000 & 0.885 & 0.000 & 0.000 & 0.105 & 0.010 & 0.000 & 0.000 & 3.098$\,10^{-4}$\\
$\sigma\bm{k}_7$ &
 & $\pm$0.345 & & & $\pm$1.029 & $\pm$0.684 & &  \\
$\sigma'\bm{k}_7$ &
 & $\pm$0.361 & & & $\pm$1.077 & $\pm$0.717 & &  \\[0.8ex]

$\bm{k}_{8}$ &
 0.366 & 0.000 & 0.000 & 0.000 & 0.231 & 0.000 & 0.403 & 0.000 & 3.162$\,10^{-4}$\\
$\sigma\bm{k}_8$ &
$\pm$0.664 & & & & $\pm$0.042& &$\pm$0.633 & \\
$\sigma'\bm{k}_8$ &
$\pm$0.680 & & & &$\pm$0.043 & &$\pm$0.648 & \\[0.8ex]

$\bm{k}_{9}$ &
 0.395 & 0.000 & 0.589 & 0.000 & 0.017 & 0.000 & 0.000 & 0.000 & 3.261$\,10^{-4}$\\
$\sigma\bm{k}_9$ &
$\pm$0.520 & &$\pm$0.773 & &$\pm$0.255 & & & \\
$\sigma'\bm{k}_9$ &
$\pm$0.535 & &$\pm$0.795 & &$\pm$0.262 & & & \\[0.8ex]

$\bm{k}_{10}$ &
 0.000 & 0.824 & 0.116 & 0.000 & 0.000 & 0.061 & 0.000 & 0.000 & 3.769$\,10^{-4}$\\
$\sigma\bm{k}_{10}$ &
 &$\pm$0.944 &$\pm$1.133 & & & $\pm$0.190 & & \\
$\sigma'\bm{k}_{10}$ &
 &$\pm$0.983 &$\pm$1.179 & & & $\pm$0.198 & & \\[0.8ex]

$\bm{k}_{11}$ &
 0.546 & 0.000 & 0.000 & 0.268 & 0.000 & 0.000 & 0.000 & 0.186 & 3.986$\,10^{-4}$\\
$\sigma\bm{k}_{11}$ &
$\pm$0.308 & & & $\pm$0.180 & & & &$\pm$0.480 \\
$\sigma'\bm{k}_{11}$ &
$\pm$0.315 & & & $\pm$0.183 & & & &$\pm$0.490 \\[0.8ex]

$\bm{k}_{12}$ &
 0.428 & 0.000 & 0.000 & 0.352 & 0.000 & 0.000 & 0.220 & 0.000 & 5.458$\,10^{-4}$\\
$\sigma\bm{k}_{12}$ &
$\pm$0.609 & & &$\pm$0.060 & & &$\pm$0.564 & \\
$\sigma'\bm{k}_{12}$ &
$\pm$0.621 & & &$\pm$0.061 & & &$\pm$0.576 & \\[0.8ex]

$\bm{k}_{13}$ &
 0.014 & 0.864 & 0.000 & 0.000 & 0.122 & 0.000 & 0.000 & 0.000 & 6.746$\,10^{-4}$\\
$\sigma\bm{k}_{13}$ &
$\pm$1.010 &$\pm$1.124 & & &$\pm$0.117 & & & \\
$\sigma'\bm{k}_{13}$ &
$\pm$1.055 &$\pm$1.174 & & &$\pm$0.122 & & & \\[0.8ex]

$\bm{k}_{14}$ &
 0.000 & 0.756 & 0.000 & 0.000 & 0.000 & 0.097 & 0.147 & 0.000 & 8.087$\,10^{-4}$\\
$\sigma\bm{k}_{14}$ &
 &$\pm$1.714 & & & & $\pm$0.178 &$\pm$1.538 &  \\
$\sigma'\bm{k}_{14}$ &
 &$\pm$1.771 & & & & $\pm$0.183 &$\pm$1.589 & \\[0.8ex]

$\bm{k}_{15}$ &
 0.401 & 0.000 & 0.556 & 0.044 & 0.000 & 0.000 & 0.000 & 0.000 & 2.269$\,10^{-3}$\\
$\sigma\bm{k}_{15}$ &
$\pm$0.611 & &$\pm$1.281 & $\pm$0.671& & & & \\
$\sigma'\bm{k}_{15}$ &
$\pm$0.627 & &$\pm$1.315 &$\pm$0.689 & & & & \\[0.8ex]

$\bm{k}_{16}$ &
 0.162 & 0.601 & 0.000 & 0.236 & 0.000 & 0.000 & 0.000 & 0.000 & 2.604$\,10^{-3}$\\
$\sigma\bm{k}_{16}$ &
$\pm$1.154 &$\pm$1.379 & &$\pm$0.228 & & & & \\
$\sigma'\bm{k}_{16}$ &
$\pm$1.184 &$\pm$1.414 & &$\pm$0.234 & & & & \\
\hline
\end{tabular}